\documentclass[12pt]{article}
\newcommand{\ra}{\rightarrow}

\newcommand{\be}{\begin{equation}}
\newcommand{\ee}{\end{equation}}
\newcommand{\ba}{\begin{eqnarray}}
\newcommand{\ea}{\end{eqnarray}}
\newcommand{\bi}{\begin{itemize}}
\newcommand{\ei}{\end{itemize}}

\newcommand{\Z}{{\bf Z}}

\newcommand{\nn}{\nonumber}

\usepackage{amsfonts}
\usepackage{amsmath, amssymb}
\usepackage{graphicx}
\usepackage{psfrag}
\begin{document}
\baselineskip=15.5pt
\renewcommand{\theequation}{\arabic{section}.\arabic{equation}}
\pagestyle{plain} \setcounter{page}{1}
\begin{titlepage}

\leftline{\tt hep-th/0404101}

\vskip -.8cm

\rightline{\small{\tt CALT-68-2491}}

\begin{center}

\vskip 3.7 cm

{\Large {\bf D Branes and Phases on String Worldsheet}}

\vskip 2.5cm
{\large Takuya Okuda
and Hirosi Ooguri}

\vskip 0.8cm

California Institute of Technology, Pasadena, CA 91125, USA

\smallskip

\vskip 3.5cm

 {\bf Abstract}

\end{center}

We generalize the worldsheet derivation of the topological
open/closed string duality given in hep-th/0205297 to cases 
when there are different types of D branes on the open string 
side. We use the mirror Landau-Ginzburg description to clarify 
the correspondence between D branes on the open string side 
and $C$ phases on the closed string side. We also discuss the
duality from the point of view of the B model.


\end{titlepage}

\newpage


\section{Introduction}

The large $N$ duality between open and close string theories
has played important roles in recent development in string theory.
The AdS/CFT correspondence \cite{Maldacena:1997re} has provided insights into
quantum gravity and strongly coupled gauge theories. The topological
open/closed string duality \cite{Gopakumar:1998ki} has uncovered relations
between spectral density of random matrix models
and geometry of Calabi-Yau manifolds and lead to the construction 
of matrix models to compute effective superpotential terms 
for supersymmetric gauge theories in four dimensions \cite{Dijkgraaf:2002dh}. 

In the case of the topological
string duality, the string coupling constant $g_s$ is the same for both
open and closed string sides, and the 't Hooft couplings of the open
string are identified with geometric moduli of the closed string side.
Thus, it should be possible to give a microscopic explanation of the
duality order by order in the string coupling expansion, namely
from the point of view of the string worldsheet. (This should also
be the case for the AdS/CFT correspondence when the coupling constant
does not run.) In \cite{Ooguri:2002gx},
the worldsheet derivation
was given for the case when the open string side is defined on
the cotangent space of a three-sphere $T^*S^3$ with $N$ D brane
wrapping  the base $S^3$. The closed string side is the
resolved conifold whose K\"ahler modulus $t$, which is the size 
of the blown up $S^2$, is identified with the 't Hooft coupling
$g_s N$  of the open string side. The strategy was to start with
the closed string side and expand string amplitudes for small $t$.
It has been noted in \cite{Witten:1993yc,Witten:1995zh} 
that in the linear sigma model  
description  the closed string worldsheet at $t=0$ develops a new 
non-geometric phase -- the $C$ phase -- besides
the geometric phase (Higgs phase) which flows to the non-linear 
sigma model for the resolved conifold in the IR limit. 
In \cite{Ooguri:2002gx}, 
it was shown that the $C$ phase can contribute to closed string 
amplitudes only in the following two cases:

(1) Each domain in the $C$ phase has the topology
of a disk.

(2) The entire worldsheet is in the $C$ phase. 

\noindent
Moreover it was found that
each disk in the $C$ phase contributes a factor of $t$
to the amplitudes. By interpreting $C$ domains as holes on the
worldsheet, the sum over disks in the $C$ phase reproduces the 
open string Feynman diagram expansion with $t$ being identified with the 
't Hooft coupling. On the other hand, worldsheets that are
entirely in the $C$ phase do not correspond to any open string
diagrams. It was shown that the sum of such worldsheets to all
order in the string coupling constant expansion correctly 
captures the gauge group volume ${\rm vol}\ U(N)$ for the open 
string, which indeed does not come from Feynman diagrams
but is needed to reproduce open string amplitudes.

Thus, this approach shows that the closed string on the resolved
conifold has the open string expansion with the K\"ahler modulus
$t$ being equal to the 't Hooft coupling on the open string side.
One may be more ambitious and try to reproduce the D brane
boundary condition. Some of the basic features of the
boundary condition can been seen in this approach: at the 
interface of the $C$ phase and
the geometric phase, one finds that the worldsheet is pulled toward
the apex of the conifold where the $S^2$ shrinks to
zero size at $t=0$ and the $S^3$ emerges after the conifold 
transition. However the precise boundary condition
for the D branes wrapping  the $S^3$ -- for example, the
Neumann boundary condition along the branes -- has not been reproduced
in this approach since the linear sigma
model does not describe the geometry with $S^3$ of finite size.
Although the size of $S^3$, being a complex structure modulus, 
is not relevant in the A-model and can be infinitesimal,
it is desirable to understanding how D brane
boundary conditions are reproduced from the closed string 
point of view.

In this paper, we will  clarify the relation between D branes
in open string and $C$ phases in closed string by studying
cases in which there are several D branes on the open string side.
We will consider the open string on the $\Z_p$ quotient of 
$T^* S^3$, whose base is the lens space $S^3/\Z_p$ \cite{Aganagic:2002wv}. 
Since
the fundamental group of $S^3/\Z_p$ is $\Z_p$, there are $p$
different types of D branes whose holonomies along the homotopy
generator are 
given by $e^{2\pi i a/p}$ ($a=0,1,..., p-1$). It turns
out that the closed string dual has $p$ different $C$ phases
on the worldsheet, and we show how each $C$ phase is identified
with the corresponding D brane by studying the behavior of linear 
sigma model variables at the interface of each $C$ phase with 
the geometric phase. We find it is useful to use the Landau-Ginzburg
B model which is $T$-dual to the linear sigma model \cite{Hori:2000kt}
since much of the analysis in the B-model can be carried out
at the classical level. We will also discuss how this can be seen
from the point of view of the mirror manifold. 

It is straightforward to apply the worldsheet
derivation in \cite{Ooguri:2002gx} and in this paper
to other toric Calabi-Yau manifolds which are known
to have open string duals \cite{Aganagic:2001ug},
for example the one described by 
$Q_1=(1,-1,0,1,-1)$ and $Q_2=(0,0,1,-2,1)$.
It would be interesting to analyze
closed string theories on more general toric manifolds and to
discover new large $N$ dualities.

This paper is organized as follows. In section 2, we will
review the worldsheet derivation of \cite{Ooguri:2002gx} in the
conifold case. We will use the Landau-Ginzburg B-model to
simplify some of the steps in the original derivation.
We will then extend the analysis to the quotients
of the conifold and explain the correspondence
between D branes and $C$ phases from the Landau-Ginzburg
description. In section 3, we will discuss the open/closed
string duality as seen from the point of view of the mirror
manifolds. 

\section{D branes and phases}\label{toric}
\subsection{Review of the conifold case}

In \cite{Gopakumar:1998ki}, it was conjectured that the A-type topological 
closed string theory on the resolved conifold
with the K\"ahler modulus $t$ is equivalent to the A-type
topological open string theory on the cotangent
space $T^* S^3$ (or equivalently on the
deformed conifold) with $N$ D branes wrapping  the base
$S^3$. Here the K\"ahler modulus $t$
of the closed string side is identified with the 't Hooft
coupling $g_s N$ of the open string side, with the string
coupling constant $g_s$ being the same on both sides. 
There have been several nontrivial checks of the conjecture
\cite{Ooguri:1999bv,Labastida:2000zp,Ramadevi:2000gq,Labastida:2000yw}. 
Finally, a worldsheet derivation
of the open/closed string 
duality was given in \cite{Ooguri:2002gx}. In this subsection, 
we will review the derivation, 
using the mirror Landau-Ginzburg model \cite{Hori:2000kt} to simplify
some of the steps.

The strategy is to start with the closed string side,
expand string amplitudes around $t=0$, and show that 
a sum over open string Feynman diagrams emerges
in the $t$-expansion. Since $t=0$ is a singular 
limit of the target space, it is useful to describe the 
worldsheet by the linear sigma model. The
worldsheet theory consists of 
four chiral superfields $\Phi_1,\Phi_2, \Phi_3, \Phi_4$ 
with charges given by
\be  Q=(1,1,-1,-1), \ee
and a gauge multiplet coupled to $Q$. Following \cite{Witten:1993yc},
we rearrange the gauge multiplet into a twisted chiral 
multiplet $\Sigma$.
When $t$ is non-zero, there is a potential for the
scalar field $\sigma$ which is the lowest component of $\Sigma$
and the linear sigma model flows
to the non-linear sigma model with the resolved conifold
as the target space. In this description, the singularity 
at $t=0$ is characterized by the fact that the potential
for $\sigma$ disappears in the limit and it can 
become indefinitely large without costing the worldsheet
action -- a new non-compact and non-geometric phase emerges 
on the worldsheet \cite{Witten:1993yc}. 
The idea of the derivation of the large 
$N$ duality, advocated in \cite{Gopakumar:1998ki} and quantified in \cite{Ooguri:2002gx}, is to regard domains in this new phase as holes 
on the worldsheet and
phase boundaries as representing D branes. 

The derivation of \cite{Ooguri:2002gx} can be streamlined by using 
the mirror of the linear sigma model, which is found
by performing the $T$-dual transformation on phase rotations
of $\Phi_{j}$, keeping $\Sigma$ as a spectator \cite{Hori:2000kt}. 
The $T$-dual of $\Phi_j$ are twisted chiral superfields $Y_j$ 
with periodicity $Y_j \sim Y_j + 2\pi i$. Combined with 
$\Sigma$, the mirror is a Landau-Ginzburg  model with
the superpotential $W$ given by
\ba
W=\Sigma(Y_1+Y_2-Y_3-Y_4-t)-\sum_{j=1}^4 e^{-Y_j}.
\ea
Note that, since the original model is A-twisted, the
mirror Landau-Ginzburg model is B-twisted. 
It is easy to check that, when $t$ is not equal to zero, 
there is no flat direction for the superpotential. 
It is believed that the Landau-Ginzburg model in this case  flows to the
mirror of the sigma model for the conifold. At $t=0$,
the superpotential remains flat for
\ba
&&y_1=y_2=-\log\sigma + \pi i,\nonumber\\
&&y_3=y_4=-\log\sigma \label{conifold critical},
\ea
with $\sigma$ being arbitrary. Here $y_i$ is the lowest
component scalar field in $Y_i$. Following \cite{Ooguri:2002gx},
we call this flat direction the $C$ 
phase.\footnote{This is called the $C$ phase in order to 
distinguish it from the Coulomb phase of the model,
which is decoupled from the geometric phase in the 
IR limit. For a more detailed specification 
of the $C$ phase, see \cite{Ooguri:2002gx}.} 

In \cite{Ooguri:2002gx}, it was argued that the $C$ phase is
described by a Landau-Ginzburg model for a single variable $X$ 
with an effective  superpotential 
\be W_{eff} = -{t\over X} \label{effectivesuperpotential}. 
\ee
It is straightforward to derive
this from the B-model. When $\sigma$ is large,
the potential for $Y_i$ becomes steep and we can 
integrate them out using the Gaussian approximation. 
Since the mirror Landau-Ginzburg model is B-twisted,
it is sufficient to consider an integral over
constant maps,
\be
   \int dY_1 dY_2 dY_3 dY_4 \exp(-W)
 \sim {1\over \Sigma^2} \exp(t\Sigma).
\ee
The pre-factor $1/\Sigma^2$ gives a non-canonical
measure for $\Sigma$. We can absorb it by changing
the variable $\Sigma \rightarrow X = \Sigma^{-1}$.
This gives the effective superpotential (\ref{effectivesuperpotential})
for the canonically normalized variable $X$.

We found that there are two phases in the worldsheet
theory: the geometric phase, which flows in the IR 
to the non-linear sigma model on the conifold, and the $C$ 
phase, which is non-geometric 
and is described by the effective Landau-Ginzburg model with the 
superpotential $W_{eff} = -t/X$. The functional
integral then includes a sum over domains in the $C$ phase
 on the worldsheet. Let us state the
following two facts that were shown in  \cite{Ooguri:2002gx}:

\bigskip

\noindent
(1) A domain in the $C$ phase contributes to a topological string 
amplitude only if (a) the domain has the topology of a disk
or (b) the entire worldsheet is in $C$ phase.

\medskip

\noindent
(2) Each $C$ domain of the disk topology contributes to the
amplitude by a factor of $t$. 
This follows from the  integral
\be
  \oint dX e^{t/X} = 2\pi i t. 
\ee

\medskip

\noindent
Thus, if we regard each $C$ domain of the disk topology
as a hole on the worldsheet, the closed 
topological string amplitude is expressed as a sum of open 
topological amplitudes with a boundary of each hole being 
weighted with a factor of $t$. 

When the entire worldsheet is in the $C$ phase, the closed
topological string amplitude at genus $g$ is given by
$\chi({\cal M}_g)/t^{2g-2}$, where $\chi({\cal M}_g)$
is the Euler characteristic of the moduli space of genus $g$
surfaces and is equal to $B_{2g}/2g(2g-2)$. The negative
power in $t$ (for $g \geq 2$) reflects the singularity at
$t \rightarrow 0$. Summing this up over all genera, one
obtains \cite{Ooguri:2002gx}
\be
  \sum_g \chi({\cal M}_g) \left(\frac{g_s}{ t}\right)^{2g-2}
= \sum_g {B_{2g} \over 2g (2g-2) N^{2g-2}}
= \log {\rm vol} \ U(N),
\ee
where we used $t=g_s N$. Thus, the sum over the worldsheet
in the pure $C$ phase gives the gauge volume factor ${\rm vol} \ U(N)$
for the gauge theory. 

This establishes that the closed topological string theory on the resolved
conifold is equivalent to some open topological string theory, with
gauge group $U(N)$ and 't Hooft coupling $t$. We have not yet
shown on which D branes open strings are ending. According to 
the conjecture \cite{Gopakumar:1998ki}, the D branes should be wrapping  the 
base $S^3$ of the deformed conifold. To see how the boundary
condition emerges from the closed string dual, we first note that
the transition from the resolved conifold to the deformed conifold 
is a local operation near the conifold singularity. Thus, away
from the base $S^3$, we can approximate the deformed conifold
by the geometric phase (Higgs phase) 
of the linear sigma model for the resolved conifold. Since A-model amplitudes
are independent of the complex structure, we can make the size
of $S^3$ as small as we like, making the approximation increasingly
accurate. In the $C$ phase, all the $\Phi^i$ fields 
in the sigma model become massive. Thus, $\Phi^i \rightarrow 0$
as we approach the ``hole'' on the closed string worldsheet, and
it is roughly where the base $S^3$ is located. Reproducing the
precise boundary condition for D branes wrapping  $S^3$ 
-- for example, deriving the Neumann boundary condition along
$S^3$ -- is difficult in this approach
since the linear sigma model does not describe the
geometry of the deformed conifold with finite $S^3$. 

Although reproducing a precise boundary condition for each
D brane may be difficult, one may ask if we can distinguish
different types of D branes in this approach. In the following
subsections, we will demonstrate that it is possible.

\subsection{Gauge theory on $S^3/\Z_2$}

As the first example in which there are more than one types
of D branes, we consider the Chern-Simons gauge theory on the
lens space $S^3/\Z_2$. 
Classical solutions are flat connections,
which are labeled by holonomy matrices for the homotopy generator 
of the space.
Since the fundamental group 
is $\Z_2$ in this case, 
the $U(N)$ gauge theory can have a holonomy matrix
with $N_1$ eigenvalues being $(+1)$ and $N_2$ eigenvalues being
$(-1)$, where $N=N_1+N_2$. This breaks the gauge group to $U(N_1)
\times U(N_2)$. This can be realized by considering the
topological string theory on $T^*S^3/\Z_2$, with $N_1$
D branes wrapping  the lens space with the trivial
bundle and $N_2$ D branes wrapping  the same space with
the bundle twisted by $(-1)$. 

According to the conjecture in \cite{Aganagic:2002wv}, 
the target space of the
closed string dual is the $\Z_2$ quotient of the resolved conifold. 
This space has two K\"ahler moduli, which are naturally
identified with the two 't Hooft couplings $g_s N_1$ and
$g_s N_2$ of the open string. This conjecture has also been
tested in nontrivial ways \cite{Aganagic:2002wv,Halmagyi:2003ze,
Halmagyi:2003mm}. 

The closed string worldsheet is described by a
linear sigma model with five chiral multiplets
$\Phi_i, i=0,...,4$ with two sets of charges 
\ba
Q_1&=&(-2,1,1,0,0),\nonumber\\
Q_2&=&(-2,0,0,1,1),\label{chargechoice}
\ea
and two gauge multiplets coupled to these charges.
Since $Q_1 - Q_2 = (0,1,1,-1,-1)$, solving the D term
constraint and dividing by the $U(1)$ gauge symmetry  
coupled to this combination of charges reduce 
$\phi_1,...,\phi_4$ to the resolved conifold. 
In the cone $r_1<0, r_1-r_2<0$,
since $\phi_0 \neq 0$, we can use the remaining
$Q_1$ gauge symmetry to fix the phase of $\phi_0$. This
leaves out a residual $\Z_2$ gauge symmetry acting as
\be
 (\phi_0,\phi_1,\phi_2,\phi_3,\phi_4)
\rightarrow 
 (\phi_0,-\phi_1,-\phi_2,\phi_3,\phi_4).
\ee
Thus we find the $\Z_2$ quotient of the resolved conifold
as the target space. 

 As in the
previous subsection, we rearrange the gauge
multiplets into twisted chiral superfields, $\Sigma_1$
and $\Sigma_2$, and perform the $T$-dual transformation
along phase rotations of the five chiral superfields to
arrive at the B-twisted Landau-Ginzburg model with the superpotential,
\be
W=\Sigma_1(-2Y_0+Y_1+Y_2-t_1)
+\Sigma_2(-2Y_0+Y_3+Y_4-t_2)-\sum_{i=0}^5 e^{-Y_i},\label{Wfirst}
\ee
where the two K\"ahler moduli, $t_1$ and $t_2$, are linearly
coupled to the gauge multiplets. 

Let us examine when this superpotential has flat directions.
Solving $\partial W/\partial y_i=0$, we find
\ba
&&y_0  =  -\log(2\sigma_1+2\sigma_2), \nonumber\\
&&y_1  =  y_2 = -\log(-\sigma_1), \nonumber\\
&&y_3 = y_4 = -\log(-\sigma_2).\label{ysol}
\ea
By substituting this into the remaining equations,
\be
  \frac{\partial W}{\partial \sigma_a}
= -2 y_0 + y_{a} + y_{a+1} - t_a = 0 ~~(a=1,2),
\ee
we find
\ba
e^{t_1} &=&4\left(1+\frac{\sigma_2}{\sigma_1}\right)^2
\nonumber\\
e^{t_2}&=&4\left(1+\frac{\sigma_1}{\sigma_2}\right)^2.
\label{whatt}
\ea
Since both $t_1$ and $t_2$ depend only on the ratio
$\sigma_1/\sigma_2$, this is possible only 
if they satisfy the relation
\be   
 \Delta = 16(e^{-t_1}-e^{-t_2})^2 -8(e^{-t_1}+e^{-t_2})
+1 = 0, \label{singularlocus}
\ee
which is obtained by eliminating $\sigma_1/\sigma_2$
from the two equations in (\ref{whatt}).  
The subspace of the K\"ahler moduli space where $\Delta=0$
is known as the singular locus\footnote{
The singular locus can also be derived from the
linear sigma model point of view \cite{Morrison:1994fr}.
In this case, we have to take into account quantum
corrections in the linear sigma model. This is in
contrast to the mirror Landau-Ginzburg description,
where the singular locus (\ref{singularlocus})
is derived from the classical analysis of the 
superpotential  (\ref{Wfirst}).}. 
If the K\"ahler moduli 
satisfy $\Delta=0$, the superpotential has a flat direction
corresponding to the scaling of $\sigma_1$ and $\sigma_2$
while keeping their ratio fixed. 

This model has two different $C$ phases. For a generic
point on the singular locus, only one of the two $C$ phases
emerges. But there is a particular point where both co-exist.
Let us consider the limit\footnote{This limit is motivated by
the fact that the two $C$ phases co-exist as we will show below.
A geometric motivation for the limit will be made clear
in section 3.}  
\be t_1, t_2 \rightarrow -\infty ,~~t_1-t_2 \rightarrow 0. \label{limit}
\ee 
In this limit, the condition (\ref{singularlocus}) for the 
singular locus gives
\be t_1 - t_2 = \pm e^{t_1/2} + O(e^{t_1}) . \label{twosolutions}\ee
For such $t_1, t_2$, we can solve (\ref{whatt}) as 
\be
 \sigma_2 = - \sigma_1  \pm e^{t_1/2}\sigma_1 + O(e^{t_1}). 
\ee
Substituting this into (\ref{ysol}), we find two flat directions
\ba
C_+ \ {\rm phase}:&& \sigma_1 = -\sigma_2 = \sigma, \nonumber \\
&& y_0 =  -\frac{t_1}{2} -\log\sigma ,\nonumber\\
&&y_1  = y_2 = -\log\sigma ,\nonumber\\
&&y_3=y_4 = -\log\sigma + \pi i. \label{phaseone} \\
&&\nonumber \\
C_-\ {\rm phase}:&& \sigma_1 = -\sigma_2 = \sigma, \nonumber \\
&& y_0 =  -\frac{t_1}{2} -\log\sigma + \pi i,\nonumber\\
&&y_1  = y_2 = -\log\sigma,\nonumber\\
&&y_3=y_4 = -\log\sigma + \pi i. \label{phasetwo} 
\ea 
Both are complex one-dimensional in the seven-dimensional
space of $(\sigma, y)$ and are parametrized 
by $\sigma$. Note that the two phases are distinguished by 
the value of $y_0$.
When $e^{t_1}$ and $e^{t_2}$ are small but finite, 
either $C_+$ or $C_-$ solves $dW=0$ depending on the 
sign $(\pm)$ on the right-hand side of (\ref{twosolutions}).
In the limit (\ref{limit}), both phases
co-exist. 

In this model, the flat coordinates $\hat t_+, \hat t_-$
are non-linear functions of the parameters $t_1, t_2$ in 
the superpotential (\ref{Wfirst}). They can be computed
either by the integrals $\int d\sigma dy e^{-W}$
with different choice of contours or by going to the mirror
of the $\Z_2$ quotient of the resolved conifold and performing period 
integrals. From the latter point of view, $\hat t_1$ and 
$\hat t_2$ are periods of two 3-cycles in the mirror manifold.
We will discuss the latter point of view in more detail
in section 3. In terms of the flat coordinates, the condition
$\Delta = 0$ is equivalent to $\hat t_+ = 0$ {\bf or} $\hat t_- =0$,
where one of the two $C$ phases emerges. 
The limit (\ref{limit}) corresponds to $\hat t_+=0$
{\bf and} $\hat t_-=0$, consistently with the fact
that both $C$ phases are realized in the limit. 

Let us examine the limit more closely. The flat coordinates
are expressed in the limit as
\ba
&&  \hat t_+ = e^{t_1/2} + t_1-t_2 + O(e^{t_1}) \nonumber \\
&& \hat t_- = -e^{t_1/2} + t_1-t_2 + O(e^{t_1}) .
\ea
Comparing this with (\ref{twosolutions}), we find that
the $C_+$ ($C_-$) phase emerges at $\hat t_+=0$ 
(at $\hat t_-=0$).
The two $C$ phases co-exist when both flat coordinates 
vanish. The two flat coordinates 
are exchanged as $(t_1,t_2) \rightarrow (t_1 + 2\pi i,
t_2+2\pi i)$ and
at the same time the two $C$ phases are also exchanged. 

Thus, at $\hat t_+ = \hat t_- = 0$, both $C$ phases
as well as the geometric phase co-exist on the worldsheet. 
We claim that the two $C$ phases correspond to the two
types of D branes with different holonomies around
the homotopy generator $\gamma$ of $S^3/Z_2$. 
This can be shown in the following three steps:

\medskip

\noindent
(1) We first note that $C_+$ and $C_-$
are distinguished by the values of $y$ as in (\ref{phaseone}) and
(\ref{phasetwo}), and that they are related to each other by a shift
of $y_0$ by $\pi i$, one half of the periodicity of $y_0$. 

\medskip

\noindent
(2) Since a shift of $y_i$ in the imaginary direction is
$T$-dual to a phase rotation of $\phi_i$ for each $i=0,...,5$,
the $C$ branches represent D branes wrapping
around the phase rotations of $\phi$'s. Since 
$C_+$ and $C_-$ differ
by the shift of one half of the period of $y_0$, 
the corresponding two types of D branes
in term of the dual $\phi$ variables
are related to each other by a multiplication
of $(-1)$ to the holonomy of the gauge field 
around a $2\pi$ phase rotation of $\phi_0$.
 
\medskip

\noindent
(3) What remains is to identify 
this $(-1)$ as the relative
holonomy around the homotopy generator $\gamma$
on the D brane worldvolume $S^3/\Z_2$. 
Since the fundamental groups of 
$S^3/\Z_2$ and $T^* S^3/\Z_2$ are isomorphic
and since the conifold transition
is a local operation near the singularity, we can lift
$\gamma$ from the base and describe it in the linear sigma model
variables. To see that $\gamma$ is homotopic to the $2\pi$ phase
rotation of $\phi_0$, we just have to note that the latter is
gauge equivalent via the $Q_1$ gauge transformation
to the $\pi$ rotation,
\be
 (\phi_0, e^{i\theta}\phi_1,e^{i\theta}\phi_2,\phi_3,\phi_4),
~~~0\leq \theta \leq \pi,
\ee
and that this path is closed because of the $\Z_2$ quotient
described in the third paragraph of this subsection.
 
\medskip
We have established that the two $C$ phases
emerge in the limit $\hat t_+, \hat t_- \rightarrow 0$.
The boundary conditions at the interface 
of the $C$ phases and the geometric phase are related to each
other by the shift of $\pi i$ of the value of $y_0$. Via the 
$T$-duality, they are mapped to boundary conditions on the 
linear sigma model variables 
related to each other by a multiplication of $(-1)$ to
the holonomies around the homotopy generator of $S^3/\Z_p$,
 $i.e.$ the two types of D branes expected in the open string 
dual.\footnote{Since the open string is in the adjoint representation
of the gauge group, only the relative holonomy of the two types
of D branes has an invariant meaning. This is $T$-dual to the
fact that only the relative value of $y_0$ in the two $C$ phases
is relevant because of the translational invariance. 
In fact gauge theory amplitudes are invariant under
exchange of $N_1$ and $N_2$, the numbers of the two types of D branes 
\cite{Aganagic:2002wv}.}

For the same reason as in the case of the conifold discussed
in \cite{Ooguri:2002gx} and reviewed in the last subsection, 
$C$ domains contribute to topological string amplitudes
only if they are of the disk topology or if they cover
the entire worldsheet. 

The large $N$ duality conjecture states that
the 't Hooft couplings for the two types of D branes are
given by the flat coordinate $\hat t_\pm$ on the closed 
string side. This can be shown as follows. 
Each of the $C$ branches is complex one-dimensional 
parametrized by $\sigma$ in (\ref{phaseone})
and (\ref{phasetwo}).
An integral in the direction transverse to $\sigma$
imposes one linear constraint on the five $Y$ fields. 
For large value of $\sigma$, remaining four $Y$ fields 
can be integrated out in the Gaussian approximation. 
As in the conifold case, this results in a superpotential
linear in $\Sigma$ with the non-canonical measure
of $1/\Sigma^2$. Using the variable $X=1/\Sigma$, one
finds an effective superpotential $ \sim 1/X$ with
the canonical measure. To find the coefficient of 
the $1/X$ potential, we note the following two
well-known facts. 

\medskip

\noindent
(1) The genus-$g$ closed string amplitude for the Landau-Ginzburg
model with the superpotential $W= t/X$ is proportional to
$ t^{2-2g}$ \cite{Ghoshal:1995wm}. 

\medskip
\noindent
(2) The singular part of the
genus-$g$ closed string amplitude for small $\hat t_\pm$ is
proportional to ${\hat t}_\pm^{2-2g}$ \cite{Vafa:1995ta}.

\medskip

\noindent
Note that both statements follow from studies on the
closed string side and do not assume the open/closed 
string duality. Comparing them, we find that 
the effective superpotential in the $C_\pm$ phase
is given by $\hat t_\pm /X$ respectively. The disk amplitude is then
computed exactly as in the conifold case, giving rise
to the factor $\hat t_\pm$ for the $C_\pm$ phase.
This is what we wanted to show. 

\subsection{Gauge theory on $S^3/\Z_p$}\label{Zp-orbifold}

It is straightforward to generalize the result in the
previous subsection to the case of the $\Z_p$ quotient 
of the conifold for $p\geq 2$.
In this case, the conjectured gauge theory dual is on
the lens space $S^3/\Z_p$ \cite{Aganagic:2002wv}. 
Since the fundamental group
of the space is $\Z_p$,
there are $p$ different types of D branes whose holonomies
around the homotopy generator 
are given by $e^{2\pi i a/p}$ ($a=0,1,..., p-1$).
We would like to see how they are identified with $p$
different $C$ phases in the closed string dual. 

The worldsheet of the closed string on the $\Z_p$ quotient
of the resolved conifold can be described by the linear
sigma model with $(p+3)$ chiral fields $\Phi_0, \Phi_1,
..., \Phi_{p+2}$ coupled to $p$ gauge fields with 
the following charge vectors \cite{Halmagyi:2003mm},
\be
\begin{array}{lrrrrrrrrrrrrrl}
&&\Phi_0,&\Phi_1,&\Phi_2,&\Phi_3,&\Phi_4,&\Phi_5,&\ldots,&\Phi_{4+j}
,&\cdots,&\Phi_{p+2}\\
&&&&&&&&&&&\\
Q_0&=(&0,&1,&1,&-1,&-1,&0,&\ldots,&0,&\ldots, &0),\\
Q_j&=(&-(j+1),&j,&0,&0,&0,&0,&\ldots,& 1,&\ldots,&0),\\
Q_{p-1}&=(&-p,&p-1,&1,&0,&0,&0,&\ldots, &0,&\ldots,&0),
\label{chargevectors}
\end{array}
\ee
where $j=1,\cdots,p-2$. Let us show that that this indeed describes
the $\Z_p$ quotient of the resolved conifold in the cone,
\ba
&&r_0<0,\nn\\
&&r_{p-1}<0,\nn\\
&&-r_j+\frac{j+1}{p}r_{p-1}<0 ~~~~ ( 1\leq j\leq p-2).
\label{zp-orbifold}
\ea
We write $U(1)_a$ for the $U(1)$ generated by $Q_a$, and 
the corresponding D term is $D_a:=\sum_i Q_{ai}|\phi_i|^2-r_a$
($a=1,...,p-1$).
If $r_0<0$, $\{D_0=0\}/U(1)_0$ describes the resolved conifold.
If $r_{p-1}<0$, $D_{p-1}=0$ does not allow $\phi_0$ to vanish.
Thus, we can use the $U(1)_{p-1}$ gauge symmetry to fix
the phase of $\phi_0$. Since $\phi_0$ carries $(-p)$ units
of $Q_{p-1}$, there is a $\Z_p$ 
residual gauge symmetry of $U(1)_{p-1}$ given by
\be
(\phi_0,\phi_1,\phi_2, \phi_{3\leq i\leq p-1})
\ra (\phi_0,e^{-\frac{2\pi i}{p}}\phi_1,
e^{\frac{2\pi i}{p}}\phi_2, \phi_{3\leq i\leq p-1}).
\ee
Other gauge groups $U(1)_{1\leq j\leq p-2}$ are completely fixed
if $-r_j+\frac{j+1}{p}r_{p-1}<0$ since
$0=D_j-\frac{j+1}{p}D_{p-1}=\left(\frac{j-p+1}{p}\right)|\phi_1|^2
-\frac{j+1}{p}|\phi_2|^2+|\phi_{4+j}|^2-r_j+\frac{j+1}{p}r_{p-1}$
requires $\phi_{4+j}$ to be non-zero. Thus, we obtain the ${\bf Z}_p$
quotient of the resolved conifold as a space of solutions to the
D term constraints 
up to gauge transformations. 

The mirror Landau-Ginzburg model has the superpotential
\ba W&=& \Sigma_0\big(Y_1+Y_2-Y_3-Y_4-t_0\big)+\sum_{j=1}^{p-2} \Sigma_j
\big(-(j+1)Y_0+jY_1+Y_{4+j}-t_j\big) \nonumber\\
&&+\Sigma_{p-1}\big(-pY_0+(p-1)Y_1+Y_2-t_{p-1}\big)
-\sum_{j=0}^{p-1}e^{-Y_j}. \ea 
 As in the previous subsection, we look for flat directions
of the potential. Solving $\partial W /\partial y_i=0$ ($i=0,\ldots,p+2$)
gives
\ba
y_0& = & -\log\left(\sum_{j=1}^{p-1} (j+1)\sigma_j\right)\nonumber\\
y_1& = & -\log\left(-\sigma_0 - \sum_{j=1}^{p-1} j\sigma_j\right)\nonumber\\
y_2&=& -\log\left(-\sigma_0 - \sigma_{p-1}\right)\nonumber\\
y_3&=&-\log \sigma_0 \nonumber\\
y_4&=& -\log\sigma_0 \nonumber \\
y_{4+i}&=&-\log(-\sigma_i),~~~~(i=1,\ldots,p-2).\label{flatnessy}
\ea
Substituting them into $\partial W/\partial \sigma_a=0$
($a=0,\ldots,p-1$) gives $p$ relations of the form,
\ba
e^{t_0} &=& \frac{\sigma_0^2}{(\sigma_0
+\sigma_{p-1})(\sigma_0+\sum_j j \sigma_j)} \nonumber \\
e^{t_k} & = & \frac{(-\sum_j (j+1)\sigma_j)^{k+1}
}{\sigma_k(\sigma_0 + \sum_j j\sigma_j)^k } \nonumber \\
e^{t_{p-1}} & = & \frac{(-\sum_j (j+1) \sigma_i)^p
}{(\sigma_0 + \sigma_{p-1})
(\sigma_0 + \sum_j j \sigma_j)^{p-1}} , \label{flatnesssigma}
\ea
where $k=1,\ldots,p-2$. Note that the right-hand sides
are functions of the ratios of $\sigma$'s.
Since there are $p$ relations for $(p-1)$ variables, 
there is no solution for generic
values of $t$ and thus no flat direction for $W$. 
The singular locus, where a flat direction
emerges, is determined by eliminating $\sigma_1/\sigma_0,\ldots,
\sigma_{p-1}/\sigma_0$
from these $p$ equations. The analysis of each of 
$C$ phase can be done as in the $\Z_2$ case. For example,
since each $C$ phase is complex one-dimensional parametrized
by the scaling of $\sigma$'s, the functional integral
over $\sigma$ in the transverse direction imposes
$(p-1)$ linear constraints on $(p+3)$ $Y$ variables,
leaving $4$ linear combinations of $Y$ free.
The functional integrals of these $4$ fields can be
done in the Gaussian approximation, giving rise to
the effective Landau-Ginzburg model with the
$1/X$ superpotential. 
  
To understand how $p$ different $C$ phases emerge,
it is useful to make the following change of variables,
\ba
Y'_0 &:=& Y_0 + {t_{p-1} \over p},\nonumber\\
Y'_2:&=&-p Y_0+(p-1)Y_1+Y_2-t_{p-1}, \nonumber\\
Y'_4:&=&-Y_1-Y_2+Y_3+Y_4+t_0,\nonumber\\
Y'_{4+j}:&=&-(j+1)Y_0+jY_1+Y_{4+j}-t_j,
 \ea
so that the superpotential takes the form,
\ba &&W=-\Sigma_0Y'_4+\sum_{j=1}^{p-2}\Sigma_j Y'_{4+j}
+\Sigma_{p-1}Y'_2
-
\big(e^{\frac{t_{p-1}}{p}} e^{-Y'_0}+e^{-Y_1}
+e^{-Y'_2-pY'_0+(p-1)Y_1}\nonumber\\
&&~~+e^{-Y_3} +e^{-Y'_4-2Y'_2-pY_0+(p-2)Y_1+Y_3+t_0}
+\sum_{j=1}^{p-2}
e^{-Y'_{4+j}+\frac{j+1}{p}t_{p-1}-(j+1)Y'_0+jY_1-t_j}\big). 
\nonumber \\
&&
\ea 
In the
limit
\ba && t_0\ra 0,\nn \\
&&t_{p-1}\ra -\infty,\nn \\
&&\frac{j+1}{p}t_{p-1}-t_j\ra-\infty, \label{limitzp}
\ea
the superpotential becomes \ba
W&=&-\Sigma_0Y'_4+\sum_{j=1}^{p-2}\Sigma_j Y'_{4+j}
+\Sigma_{p-1}Y'_2
-\big( e^{-Y_1}+e^{-Y'_2-pY'_0+(p-1)Y_1}\nonumber\\
&&+e^{-Y_3} +e^{-Y'_4-2Y'_2-pY'_0+(p-2)Y_1+Y_3} \big).
\ea
Extremizing $W$ leads to $p$ different families of solutions
\ba
C_k ~{\rm phase:}~~~
&&y'_0=-\log(-\sigma_0)+\frac{2\pi i}{p}k, \nn\\
&&y_1=-\log(-\sigma_0),~~y'_2=0\nn\\
&&y_3=-\log(\sigma_0),~~y'_4=0\nn\\
&&y'_{4+j}=0,~~\sigma_j=0~~~~~ (1\leq j\leq p-2),\nn\\
\ea
where $k=0,1,...,p-2$. We found that 
$p$ different $C$ phases co-exist in the limit 
(\ref{limitzp}). 

The $p$ different $C$ phases are related to each other as
$C_k \rightarrow C_{k+1}$ under 
the shift $y_0' \rightarrow y_0' +2\pi i/p$.
In terms of the original $y$ variables, the shift is expressed
as
\ba
&&(y_0,y_1,y_2,y_3,y_4,y_{4+j})\nn\\
&&~~~
\ra\left(y_0+\frac{2\pi i}{p} ,y_1 ,y_2,
y_3 ,y_4, y_{4+j}
+\frac{2\pi i}{p}(j+1) \right).
\label{shift}
\ea
As explained in the second paragraph of this subsection, 
the homotopy generator of $S^3/\Z_p$ is the path
\be
(\phi_0,e^{-i\theta}\phi_1,e^{i\theta}
\phi_2,\phi_3,...),~~~0 \leq \theta \leq
{2\pi \over p} .\label{path}
\ee
Note that a $2\pi$ rotation of
$\phi_1$, keeping other variables fixed, is contractible
even if one is away from the apex of the conifold since 
$\phi_1$ can vanish while maintaining the D term constraints . 
Thus, the 
$(-2\pi/p)$ rotation
of $\phi_1$ in the above can be replaced by the
$2\pi(p-1)/p$ rotation. 
Under the $U(1)_{p-1}$ gauge transformation with respect
to the charge vector $Q_{p-1}$ in (\ref{chargevectors}),
this is gauge equivalent to a $2\pi $ phase rotation
of $\phi_0$.
Since the map from $C_k$ to $C_{k+1}$ ($k=0,...,p-1$)
involves the $2\pi i/p$ shift of $y_0$, their holonomies 
under the $2\pi$ phase rotation of $\phi_0$ differ 
by $e^{2\pi i/p}$. Namely, the relative holonomy
of $C_k$ and $C_{k+1}$ 
around the homotopy generator of $S^3/\Z_p$
is equal to $e^{2\pi i/p}$,
precisely reproducing the large $N$ duality stated in the first 
paragraph of this subsection.

In principle, we can carry out the analysis further in this
approach and find the flat coordinates explicitly in the
Landau-Ginzburg description. 
However, it is more convenient and geometrically more intuitive
to use the mirror manifold, which can be obtained by partially
performing the functional integral of the Landau-Ginzburg model
and making some change of variables \cite{Hori:2000kt}. We are going
to discuss it in the next section.

\section{B-model large $N$ dualities}\label{DV}

The large $N$ duality from the point of view of the B-model
has lead to the discovery of the relation between spectral
density of matrix models and Calabi-Yau geometry. In this
section, we will give a worldsheet derivation of this and
related dualities.

\subsection{Mirror of the $\Z_p$ quotient of the resolved conifold}

In the last section, we considered 
the $\Z_p$ quotient of the resolved conifold.
Here we will study the same problem but from the
point of view of its mirror manifold.
The mirror Calabi-Yau manifold is given by
the equation \cite{Halmagyi:2003mm},
\ba
0&=&G(x_1,x_2,u,v)\nonumber\\
&\equiv& x_1^2+x_2^2+(e^v-1)(e^{pu +v}-1)
 +e^{t_0}-1-e^{t_{p-1}/p+u+v}
-\sum_{j=1}^{p-2}e^{\frac{j+1}{p}t_{p-1}-t_j+(j+1)u+v}.\nonumber\\
\ea
The non-linear sigma model on the above non-compact Calabi-Yau
can be realized as the IR limit of the Landau-Ginzburg model
with chiral superfields $\Lambda,X_1,X_2,U,V$ and the superpotential
\ba
W&=&\Lambda G(X_1,X_2,U,V),
\ea
where the scalar components of $U$ and $V$ are defined modulo $2\pi i$.
Indeed, as long as the geometry is smooth so that there
is no solution to $G=dG=0$, the only solutions to $dW=0$
are $\Lambda=G=0$.
Excitations transverse to $\Lambda=G=0$ are massive.
Hence in the low-energy the theory flows to
the non-linear sigma model on the geometry $G=0$.
Such a Landau-Ginzburg model was
considered in \cite{Witten:1995zh} in the case
of deformed conifold.
This Landau-Ginzburg model is related to the model used
in subsection \ref{Zp-orbifold}
by partially carrying out the functional integral
and by making change of variables \cite{Hori:2000kt}.
We will show that the worldsheet phase structure found
in subsection \ref{Zp-orbifold} can also be obtained
from this Landau-Ginzburg model. 

In the limit
 \ba&& t_0\ra 0,\nn\\
&& t_{p-1}\ra -\infty,\label{Zp-limit}\nn\\
&&\frac{j+1}{p}t_{p-1}-t_j\ra-\infty,
\ea
we have
\ba
G\sim x_1^2+x_2^2+(e^{pu+v}-1)(e^v-1)
\ea
and the geometry $G=0$ develops $p$ conifold singularities
at
\ba
(x_1,x_2,u,v)=(0,0,\frac{2\pi i}{p}a,0),\nn\\
~~~~~~~~~~~~~a=0,1,...,p-1.
\ea
The singularities of the geometry are reflected
in the worldsheet theory
as the appearance of new non-compact directions
in the space of zero-energy configurations.
Namely,
the worldsheet theory develops
$p$ new flat directions where $\lambda$, the lowest
component of $\Lambda$, is large
while $(x_1,x_2,u,v)$ are fixed to the locations
of conifold singularities.

We see that $\Lambda$ plays the same role as
the $\Sigma$ field and
$C$ phases can be defined
as flat directions where $\lambda$ becomes
large.
Different $C$ phases
are distinguished by the values of $u$.
In each $C$ domain, since $\lambda$ is large,
we can integrate out $x_1,x_2,u$ and $v$ by Gaussian approximation
which produces the measure $d\lambda/\lambda^2$.
Thus, the effective Landau-Ginzburg model of each of
the $C$ phases is again
with the $1/X$ superpotential. The coefficient
of the superpotential is given by the flat coordinate
$\hat t_a$ ($a=0,\ldots,p-1$) since each $C$ phase
is associated with shrinking of one of $p$ $3$-cycles.

In the language of the present subsection it is easy to
generalize the analysis of the singular locus and the 
periods.
Let us introduce new parameters $d_0, d_1,...,d_{p-1}$ as 
functions of $t_0,...,t_{p-1}$ so that $G$ is expressed as
\ba
G=x_1^2+x_2^2+(e^{pu+v}-1)(e^v-1)+d_0+\sum_{j=1}^{p-1}d_j e^{ju+v}.
\ea
The limit (\ref{Zp-limit}) is equivalent to $d_j\ra 0$ 
for $j=0,1,...,p-1$.
Let us evaluate 
$\Delta$ and the periods $\hat{t}_a$ in this limit.
Suppose $(u,v)$ are near $(2\pi i a/p,0)$ and
 write $(u,v)=(2\pi i a/p +\delta u,\delta v)$.
Assuming that $\delta u$ and $\delta v$ are of the
order ${\cal O}(d)$, we can expand $G$ to the quadratic order
in the variation and find
\ba
G\sim x_1^2+x_2^2+(p\delta u+\delta v)\delta v+d_0+\sum_{j=1}^{p-1}d_j 
(e^{\frac{2\pi i}{p}a})^j(1+j \delta u+\delta v).
\ea
Completing the squares and evaluating the constant piece in the leading
order puts $G=0$ in the form
\ba
x_1^2+x_2^2+x_3^2+x_4^2-\mu = 0,
\ea
with $\mu=\sum_{j=0}^{p-1}d_j e^{\frac{2\pi i}{p}aj}$.
Thus, we can choose the flat coordinate $\hat t_a$ as
\be
 \hat t_a \sim \sum_{j=0}^{p-1}d_j e^{\frac{2\pi i}{p}aj}.
\ee
Repeating this for all $a=0,1,...,p-1$, we find that
the discriminant $\Delta$ to the leading order in $d$ is given by
\be
\Delta\sim \hat t_0 \hat t_1 \cdots \hat t_{p-1}
\sim \prod_{a=0}^{p-1} \sum_{j=0}^{p-1} d_j e^{\frac{2\pi i}{p}aj}
=\mathop{\det}_{i,j}(d_{i-j\ {\rm mod}\ p}) .
\ee

\subsection{$x_1^2+x_2^2+x_3^2+w'(x_4)^2+f(x_4)=0$}

Essentially the same analysis applies to the geometry
\be
x_1^2+x_2^2+x_3^2+w'(x_4)^2+f(x_4)=0, \ee 
which has been extensively studied
in the context related to gauge theory/matrix model
correspondence \cite{Dijkgraaf:2002fc}. Here $w(x)=\frac{1}{n+1}x^{n+1}+...$ 
and $f$ are polynomials of degrees
$n+1$ and $n-1$, respectively.
This means that $w'(x)^2+f_{n-1}(x)=x^{2n}+...$
is an arbitrary polynomial of degree $2n$ with
unit coefficient of $x^{2n}$.
The non-linear sigma model on the geometry
can be realized as the IR limit of the Landau-Ginzburg
model with superpotential \be
W=\Lambda(X_1^2+X_2^2+X_3^2+w'(X_4)^2+f(X_4)). \ee

When all the coefficients in $f$ become small, $n$ conifold singularities appear and
$n$ new branches develop.
The $i$-th one is characterized by large values of $\lambda$ and
$x_1=x_2=x_3=0, x_4=a_i$ where $w'(x)=\prod_{a=1}^n(x-a_i)$.
We define the $i$-th Coulomb domain to be the place
where the scalar field $\lambda$ in
 the lowest component of $\Lambda$ becomes large,
and $x_4$ is frozen to $a_i$.
In this case,
the $i$-th Coulomb domain is described by the Landau-Ginzburg
model with $W=s_i/X$, where $s_i=\int_{A_i}\Omega=\int_{A_i}
dx_1dx_2dx_3dx_4/dG$ is
the flat coordinate. $A_i$ is the $S^3$ obtained by deforming the
conifold singularity at $x_4=a_i$.
Applying the usual arguments, 
we get the `t Hooft expansion for the B-model open string
on the blow-up of $x_1^2+x_2^2+x_3^2+w'(x_4)^2=0$.
Hence this B-model open string is large $N$ dual to
the B-model closed string on $G=x_1^2+x_2^2+x_3^2
+w'(x_4)^2+f(x_4)=0$.
This proves the large $N$ duality of the type
used in \cite{Dijkgraaf:2002fc}.

We find it interesting to look at the deformed conifold case
from this point of view.
In this case, $w(x)=x^2/2$ and the Landau-Ginzburg superpotential is \ba
W=\Lambda(X_1^2+X_2^2+X_3^2+X_4^2-\mu). \ea 
In the previous examples, the effective superpotential
$W_{eff} \sim 1/X$ are given by applying 
the Gaussian approximation to the chiral multiplet fields, 
which is valid when $|\lambda| \gg 1$. 
In the case of the deformed conifold, this approximation 
is exact for any $\lambda$ since $W$ is already quadratic in $X_1,...,X_4$. 
Thus, we obtain $W_{eff} = \mu/X$ without any 
approximation.\footnote{The point
arose from discussion with Donal O'Connell.} 
Therefore the sigma model on the
deformed conifold is equivalent to the Landau-Ginzburg
model with the $1/X$ superpotential \cite{Ghoshal:1995wm,
Mukhi}.

Since the $1/X$ Landau-Ginzburg model -- effective theory
in the $C$ phase --- is equivalent to the sigma model for 
the deformed conifold, full topological string amplitudes 
in this case are computable with worldsheets in pure $C$ 
phase alone. This means that, from the point of view of 
the open string dual, perturbative open string Feynman diagrams 
should not contribute to topological string amplitudes. Indeed,
this is consistent with the fact that the corresponding
hermitian matrix model is Gaussian and there are no perturbative
contributions \cite{Dijkgraaf:2002fc}.
This is a non-trivial check of our worldsheet analysis.

  \section*{Acknowledgments}
      
We would like to thank Mina Aganagic, Jaume Gomis, Nick Halmagyi,
Kentaro Hori, Amir Kashani-Poor, Yi Li, Marcos Mari\~no, 
Donal O'Connell, Christian Roemelsberger, and Cumrun Vafa
for useful discussions.
This research is supported in part by
DOE grant DE-FG03-92-ER40701.


\renewcommand{\baselinestretch}{0.87}
\begingroup\raggedright
\endgroup
\end{document}